\documentclass[reprint,superscriptaddress,amsmath,amssymb,floatfix,aps,prd,nofootinbib]{revtex4-1}
\usepackage{bm}
\usepackage{epsfig}
\usepackage{graphicx,epsfig}
\usepackage{hyperref}
\usepackage{ifthen}
\usepackage [english]{babel}
\usepackage{xstring}
\usepackage[applemac]{inputenc}
\usepackage{amsmath,amssymb}
\usepackage{color}
\usepackage{epstopdf}
\usepackage{caption}
\newcommand{\jcap}{JCAP}

\newcommand{\araa}{ARA\&A}

\newcommand{\aap}{Astron. Astrophys.}

\newcommand*\aj{AJ}

\newcommand{\nver}{\hat{\mathbf{n}}}

\def\be{\begin{equation}}
\def\ee{\end{equation}}
\def\ben{\begin{equation*}}
\def\een{\end{equation*}}

\def\ba{\begin{eqnarray}}
\def\ea{\end{eqnarray}}
\def\ban{\begin{eqnarray*}}
\def\ean{\end{eqnarray*}}

\newcommand{\refsec}[1]{section~\ref{sec:#1}}
\newcommand{\reftab}[1]{Tab.~\ref{tab:#1}}
\newcommand{\refeq}[1]{Eq.~(\ref{eqn:#1})}

\newcommand{\reffig}[1]{Fig.~\ref{fig:#1}}

\definecolor{darkgreen}{cmyk}{0.85,0.2,1.00,0.2}
\definecolor{purple}{cmyk}{0.5,1.0,0,0}

\usepackage{hyperref}
\usepackage{amsmath}
\usepackage{natbib}

\def\ublu{\bl'}

\newcommand{\comment}[1]{{}}
\def\beq{\begin{equation}}
\def\eeq{\end{equation}}
\def\beqn{\begin{eqnarray}}
\def\eeqn{\end{eqnarray}}

\def\cl{C_{\ell}}

\def\ba{\bm{\alpha}}

\def\2gcm{\textrm{g cm$^{-2}$}}

\def\H0{\ensuremath{\mathrm{H}_0}}

\def\bl{\bmm{l}}

\def\fsky{f_{\mathrm{sky}}}

\newcommand{\bmm}[1]{{\mathbf{#1}}}

\begin{document}

\title{Future cosmic microwave background delensing with galaxy surveys}
\author{A. Manzotti}
\email{manzotti.alessandro@gmail.com}
\affiliation{Kavli Institute for Cosmological Physics, University of Chicago, 5640 South Ellis Avenue, Chicago, IL 60637, USA}
\affiliation{Department of Astronomy and Astrophysics, University of Chicago, 5640 South Ellis Avenue, Chicago, IL 60637, USA}
\date{\today}

\begin{abstract}

The primordial B-modes component of the cosmic microwave background (CMB) polarization is a promising experimental dataset to probe the inflationary paradigm. B-modes are indeed a direct consequence of the presence of gravitational waves in the early universe. However, several secondary effects in the low redshift universe will produce \textit{non-primordial} B-modes. In particular, the gravitational interactions of CMB photons with large-scale structures will distort the primordial E-modes, adding a lensing B-mode component to the primordial signal. Removing the lensing component ("delensing") will then be necessary to constrain the amplitude of the primordial gravitational waves. Here we examine the role of current and future large-scale structure surveys in a multi-tracers approach to CMB delensing. We find that, in general, galaxy surveys should be split into tomographic bins as this can increase the reduction of lensing B-modes by $\sim 25\%$ in power in the most futuristic case. 
Ongoing or recently completed CMB experiments (CMB-S2) will particularly benefit from large-scale structure tracers that, once properly combined, will have a better performance than a CMB internal reconstruction. With the decrease of instrumental noise, the lensing B-modes power removed using CMB internal reconstruction alone will rapidly increase. Nevertheless, optical galaxy surveys will still play an important role even for CMB S4. In particular, an LSST-like survey can a achieve a delensing performance comparable to a 3G CMB experiment but with entirely different systematics. This redundancy will be essential to demonstrate the robustness against systematics of an eventual detection of primordial B-modes.

\end{abstract}

\vspace{1cm}

\maketitle

\section{Introduction}
\label{sec:intro}
In the standard cosmological paradigm the early Universe underwent a period of near-exponential expansion called ``cosmic inflation''.
All the cosmological observations agree with this picture, making it a compelling and elegant description of the Universe initial conditions. Despite the experimental effort, other possible explanations are still valid, and a conclusive evidence of inflation is still to be found.
Inflation generically predicts a stochastic background of gravitational waves \cite[see e.g.][for a review]{kamionkowski:2016}. This prediction sets inflation apart from other theories and a detection of primordial gravitational waves could be the compelling evidence cosmologists are looking for.
These primordial gravitational waves in the early Universe would imprint a unique signature on the polarized anisotropies of the CMB. For this reason, CMB polarization is a promising dataset to understand the physics of the early Universe and ultimately test inflation.
In particular, we can decompose the CMB polarization fields in Fourier space into even-parity (divergence) and odd-parity (curl) components, referred to as ``E'' and ``B'' modes.
 In the standard scenario, the B-mode polarization is a clean probe of primordial gravitational waves, because these are the only source of B-modes at the epoch of recombination.

Because the CMB B-modes provide the cleanest known observational window into the primordial gravitational wave background, improving their measurement is a major objective of current and future CMB experiments.
Even if inflationary B-modes have not been detected yet, the target value $r \gtrsim 10^{-3}$, should be reachable shortly given the level of noise expected in future CMB experiments \citep{kamionkowski:2016,abazajian:2016}.
However, just reducing the level of noise will not be enough to reach this target.

Indeed, the observed B modes are not solely sourced by early Universe physics; they are also produced by secondary effects taking place in the late-time low-redshift Universe.
Two main effects produce non-primordial B-modes: the polarized foregrounds from the Galaxy and the effect of gravitational interactions with large scale structures (LSS).
These two components need to be treated with very different techniques, and in this work, we will focus on the latter (see \cite{remazeilles:2017b} for a review on the former).
Lensing shears the CMB polarization pattern, producing ``lensing B modes'' from CMB E modes \citep{zaldarriaga98}.
This expected component has now been measured both in cross-correlation with LSSs \citep{hanson13,bicep2-collaboration:2016,polarbear2014c,van-engelen:2015,planck2015XV} and from CMB data alone \citep{polarbear2014b,bicep2a,keisler15}.

This component acts as a source of confusion for searches of the primordial gravitational wave background.
Indeed, the contamination from lensing B modes is already at the level of the instrumental noise of current experiments \citep{bk14}. Thus, together with the experimental effort to reduce the amount of noise, the effect of the lensing component must be understood.

The optimal way to reduce the lensing contributions to the B-modes is to reconstruct a template of the actual lensing B modes (or the actual large scale structure lensing potential) on the observed part of the sky and then use it to marginalize the lensing contribution in the data in a process called ``delensing''.
We can delens the observed B-modes by combining CMB polarization data (what is \textit{lensed}) with tracers of the large scale structure (what is \textit{lensing}) to reconstruct a template of the expected lensing B-modes.
Delensing has been studied for many years \cite{knox2002, kesden2002, seljak2003, simard:2015,sherwin15,smith:2012}.
Furthermore, it has recently been performed on CMB temperature data using the cosmic infrared background (CIB) as LSS tracer \cite{larsen:2016a} and on CMB temperature and polarization data using CMB data to internally reconstruct the LSS lensing potential \cite{carron:2017}. Finally, the highest B-mode delensing efficiency has been achieved with the South Pole Telescope (SPT) and Herschel data in which $~28\%$ of the lensing power was removed \cite{manzotti:2017}.

For future experiments, we need to increase the delensing efficiency by almost a factor of 3 to fully exploit the expected instrumental capabilities \cite{abazajian:2016}.
In this paper, we propose and study a possible way: using future galaxy surveys as tracers of the lensing potential in addition to other probes such as the SKA radio continuum survey, cosmic infrared background and internal CMB reconstruction.
Furthermore, we point out how using redshift information through tomographic binning can improve the delensing efficiency of galaxy survey.
These will translate into a better reconstruction of the B-modes in the measured patch, and, as a consequence will improve the constraints on inflation through delensing. We model several actual and future surveys, and after computing the residual B-modes, we forecast the resulting statistical uncertainties on the amplitude and the shape of the inflationary tensor perturbations for ongoing or recently completed CMB experiments (S2) as well as the next generation (S3) and the planned fourth generation (S4).

We organize this article as follows: we describe the LSS tracers used in this analysis in \refsec{model}. In \refsec{th} we define the lensing B-mode component and the residual power after delensing with tracers of the lensing potential. The main result of this work is presented in \refsec{for} : the improvement of inflationary parameters constraints due to delensing with CMB and LSSs. We conclude in \refsec{concl}.

\section{Lensing potential tracers}
\label{sec:model}
In this section, we introduce the different large scale structure tracers considered in this work to reconstruct the lensing potential. We will see how large scale structures distorts primordial E-modes generating a non-primordial B-mode component in the next section.
Also, we define the power spectra that we will use later in \refsec{th}.

Large scale structure surveys usually probe the 3D matter overdensities as a 2D field projected along the line of sight:
\begin{equation}
\delta^{i}(\nver) = \int_0^{\infty} dz\, W^{i}(z)\delta(\chi(z)\nver,z).
\label{eqn:wkernel}
\end{equation}
where $\delta(\chi(z)\nver,z)$ corresponds to the dark matter overdensity field at a comoving distance $\chi(z)$ and at a redshift $z$ in the angular direction $\nver$.
Using the Limber approximation \cite{limber53} we can compute the power spectra of two large-scale structure fields $i,j$ as:
\begin{equation}
C_{\ell}^{ij}= \int_0^{\infty} \frac{dz}{c} \frac{H(z)}{\chi(z)^2}\, W^{i}(z)W^{j}(z)P(k = \ell/\chi(z),z).
\label{eqn:wkappa}{}
\end{equation}

In this equation, $H(z)$ is the Hubble factor at redshift $z$, $c$ is the speed of light, and $P(k,z)$ is the matter power spectrum evaluated at redshift $z$.
We will now describe the kernels $W^{i}(z)$ for each of the tracers used in this work.

\subsection{CMB lensing potential}
\label{sec:kappaCMB}
We start from the CMB lensing potential.
The lensing kernel $W^{\kappa}$ is:
\begin{equation}
W^{\kappa}(z) = \frac{3\Omega_{\rm m}}{2c}\frac{H_0^2}{H(z)}(1+z)\chi(z)\frac{\chi_*-\chi(z)}{\chi_*},
\end{equation}
where $\chi_*$ is the comoving distance to the last-scattering surface at $z_*\simeq 1090$, $\Omega_{\rm m}$ and $H_0$ are the present day values of the Hubble and matter density parameters, respectively. As can be seen in \reffig{kernel}, the lensing kernel $W^{\kappa}(z)$ is a broad function of redshift peaking around $z=1.8$ but extending to high redshift.

The CMB lensing potential is \textit{the} field that we need in order to reverse the effect of large scale structure and delens the CMB.
However, the lensing potential can also be reconstructed using the CMB itself. In that case, we can treat it as a noisy tracer of the true field. Both the CMB lensing field and its noisy reconstructed counterpart have the same kernel $W^{\kappa}(z)$.
However, when computing the power spectrum of the latter, we need to add a noise component. Given the instrumental noise level and the beam, we can calculate the reconstruction noise $N_{\ell}^{\kappa \kappa}$, and so
\beq\label{eqn:noisekappa}
\cl^{\kappa_\mathrm{rec} \kappa_\mathrm{rec}} = \cl^{\kappa \kappa } +N^{\kappa \kappa}_{\ell}
\eeq
In this work, the level of noise is computed assuming an iterative approach to the CMB lensing reconstruction as described in \cite{smith:2012,hirata:2003}. 
Note that, as pointed out for example in \cite{smith:2012,hirata:2003}, this iterative approach improves the CMB lensing reconstruction if compared to a quadratic estimator (QE) approach \cite{okamoto:2003,hu:2002} for S3 and S4 noise levels. For these two cases, we also compute the delensing efficiency obtained when the CMB lensing is reconstructed with the standard quadratic estimator (QE).
Furthermore, we do not deal here with the presence of an internal delensing bias. Indeed we assume that we can use all the CMB scales for lensing reconstruction even if these scales are also used to constraint inflationary parameters. These is supported by several promising approaches \cite{sehgal:2017,namikawa:2017} although they have not been applied to low noise data yet (see \cite{carron:2017} as an example where the internal delensing bias was removed).

\subsection{Galaxies}

The normalized galaxy clustering kernel is:
\begin{equation}
\label{eqn:wg}
\begin{split}
W^{g}(z) &= \frac{b(z)\frac{dN}{dz}}{\Bigl(\int dz'\,\frac{dN}{dz'}\Bigr)}.
\end{split}
\end{equation}
Here $\frac{dN}{dz}$ is the number of galaxies observed by the survey as a function of redshift while $b(z)$ is the galaxy bias that connects the amplitude of galaxy overdensities to the underlying dark matter density.
We use a linear bias independent of the angular scale, which is a reasonable assumption for the relatively large scales relevant for delensing ($\ell<1000$).
When computing the auto-spectrum of the galaxy density, a shot noise term needs to be taken into account.
To do so, we add a constant term to the power spectrum equal to the inverse of the number of galaxies per steradians.
 Different galaxy surveys in this work are then fully characterized by their $b(z)$, $\frac{dN}{dz}$ and the observed galaxy density.
We test the delensing efficiency taking into account both current surveys like the Wide-field Infrared Survey Explorer (WISE) or the Dark Energy Survey (DES) as well as future galaxy surveys like the Dark Energy Spectroscopic Instrument (DESI) and the  Large Synoptic Survey Telescope (LSST) as well as 21 cm measurement like the Square Kilometre Array (SKA).

The WISE survey observed the entire sky in the infrared \cite{wright:2010}.
We defined the redshift distribution of the WISE infrared galaxy samples following \cite{yan:2013} (see Fig. 4 therein).
To compute the noise term, we assume that the available sky after masking is around $f_{sky}=0.44$ with 50 million galaxies \cite{ferraro:2015} and that the galaxy density is approximately uniform. Furthermore we adopt a linear bias $b_{\rm{WISE}}=1.41$ obtained by cross correlating WISE with Planck lensing potential \cite{ferraro:2015}.

DES is modeled after the DES Science Verification public data release.
For DESI we used the $\frac{dN}{dz}$ in Tab. 2.3 of \cite{desi-collaboration:2016}. From that we can derive the galaxy density of 0.63 galaxies per squared arcmin.

For LSST we follow \cite{schaan:2017}: $\frac{dN}{dz}\propto z^{\alpha}\exp^{-(z/z_{0})^{\beta}}$ with $\alpha = 1.27$, $\beta = 1.02$, and $z_{0}= 0.5$.
These correspond to a median redshift for LSST galaxies around $z_{\text{median}}\sim1.1$.
Furthermore we assume a density of 26 galaxies per arcmin squared.

Finally, we consider the Square Kilometer Array (SKA).
The SKA is a planned radio array that will survey large scale structure primarily by detecting the redshifted neutral hydrogen (HI) 21cm emission line from a large number of galaxies out to high redshift.
We will assume a radio continuum survey mode where SKA will detect radio galaxies through their total emission out to very high redshift.
We model both the redshift distribution and bias of radio sources following \cite{namikawa:2016a}.
In \reffig{kernel}, we compare the CMB lensing kernel $W^{\kappa}(z)$ to the kernels $W^{g}(z)$ of all the tracers introduced here.


\subsection{Cosmic infrared Background (CIB)}
\label{sec:cib}

The CIB consists of diffuse extragalactic radiation generated by the unresolved emission from star-forming galaxies (see \cite{dole:2006} and references therein).
In these galaxies, the UV light from young stars heats the dust regions around them that then reradiates thermally in the infrared with a gray-body spectrum of $T \simeq30$K.

Following \cite{Addison:2011se}, we model the CIB power directly as $C_{\ell}^{{\rm CIB}\mbox{-}{\rm CIB}} = 3500 (l/3000)^{-1.25} {\rm Jy^2 / sr}$.
This model provides an accurate fit to several experimental results.
For the cross-spectra with the CMB lensing or other galaxy tracers, $C_{\ell}^{{\rm CIB}\mbox{-}j}$, we use the single-spectral energy distribution (SED) model of~\cite{hall:2010}.
It corresponds to the kernel:
\begin{align}\label{eq:12}
W^{\text{CIB}}(z) = b_c\ \frac{\chi^2(z)}{H(z)(1+z)^2}\ e^{-\frac{(z-z_c)^2}{2\sigma^2_z} } f_{\nu(1+z)},
\end {align}
for
\begin{equation}
f_{\nu} =
\begin{cases}
\Big( e^{\frac{h\nu}{kT}} - 1 \Big)^{-1} \nu^{\beta+3} & (\nu \leq v^{\prime}) \\ \Big( e^{\frac{h\nu^{\prime}}{kT}} - 1 \Big)^{-1} \nu^{\prime \beta+3} \Big( \frac{\nu}{\nu^{\prime}} \Big)^{-\alpha} & (\nu > v^{\prime})
\end{cases}
\end{equation}

We place the peak of the CIB emissivity at redshift $z_c = 2$ with a broad redshift kernel of width $\sigma_z = 2$ and we set $T$ = 34K and $\nu^{\prime} \approx$ 4955 GHz.
\reffig{kernel} shows how the CIB kernel peaks at higher redshift compared to other galaxy survey kernels, with a better overlap with the CMB lensing one.

There are several available CIB observations that have already been used to delens the CMB. Given its full-sky coverage and the small contribution from foregrounds contamination, a promising CIB map for future experiments is the one derived in \cite{2016A&A...596A.109P} using multi-frequency Planck data and the Generalized Needlet Internal Linear Combination (GNILC) component separation algorithm.

\section{Gravitational lensing B-mode and delensing}
\label{sec:th}
The large scale structures described in \refsec{model} have an important impact on the search of primordial CMB B-modes: they lens the primordial E-modes generating non inflationary B-modes that, then constitute an important source of noise.

Indeed the Q and U mode decompositions of the CMB photons polarization are remapped by lensing as:
\be
Q(\hat{\mathbf{n}}) = Q_{\mathrm{unlensed}}(\hat{\mathbf{n}}+\nabla\phi);~~
U(\hat{\mathbf{n}}) = U_{\mathrm{unlensed}}(\hat{\mathbf{n}}+\nabla\phi)
\ee
where the deflection angle is the gradient of the lensing potential integrated along the line of sight $\nabla\phi$.
The CMB polarization is usually decomposed into odd-parity Fourier modes E and B.
As shown in \cite{seljak:1997}, because of the symmetry of the problem, tensor perturbations are the principal source of the B-modes configuration. For this reason, primordial B-modes are a promising signature of early Universe tensor perturbations.

However, primordial B-modes are obscured by gravitational interactions between the large scale structures and the CMB that generate CMB B-modes by distorting primordial E-modes.
At first order, given the convergence field $\kappa= -\frac{1}{2}\nabla^2\phi$ introduced in \refsec{kappaCMB} the B-modes resulting from the lensing of the primordial E-modes are:
\be
B^{\mathrm{lens}}(\bl) =  \int \frac{d^2 \bl'}{(2 \pi)^2} W(\bl,\bl') E(\ublu) \kappa(\bl - \bl')
\label{eqn:blens}
\ee
where different modes contributes with a different weight:
\be
W(\bl,\bl') = \frac{2 \bl' \cdot (\bl-\bl')}{|\bl-\bl'|^2} \sin(2\varphi_{\bl,\bl'}).
\ee
Here $\varphi_{\bl,\bl'}$ is the angle between the two different modes $\bl$ and $\bl'$.
From this we get the power spectrum of the lensing component of the B-modes:
\be
C_{\ell}^{BB,\mathrm{lens}}  = \int \frac{d^2 \bl'}{(2 \pi)^2} W^2 (\bl,\bl')C^{EE}_{l'} C^{\kappa \kappa}_{|\bl-\bl'|}.
\ee

The B-mode power spectrum measured on the sky is composed of a possible primordial component $C_{\ell}^{BB,r}$ together with the lensing $C_{\ell}^{BB,\mathrm{lens}}$ contribution and the instrumental noise $N_{\ell}^{BB}$ (defined in \refeq{noise}):
\be
C_{\ell}^{BB,\mathrm{measured}} = C_{\ell}^{BB,r} + C_{\ell}^{BB,\mathrm{lens}} + N_{\ell}^{BB}.
\ee
The lensing component is a significant source of B-modes that, at large scales, corresponds to a white noise source of roughly $5 \mu K$-arcmin independent of the angular scale. This means that it is not only larger than the allowed inflationary component at scales smaller than several degrees ($r_{0.05} < 0.07$ from \cite{bicep2-collaboration:2016a}), but it is also comparable to current levels of instrumental noise.
For this reason, it is critical to characterize and eventually remove it from the data.
To do so, while other approaches are possible (\cite{larsen:2016a,carron:2017,millea:2017}), here we assume a "template approach": we build a template \refeq{blens} of the lensing B-modes in the observed patch given a measurement of the E-mode field and the lensing potential $\phi$.
While E is measured directly, we can estimate $\phi$ using "tracers" of the matter distribution that sources the potential.

\subsection{Single tracer of the lensing potential}\label{sec:single-tracer}

We will now show how the delensing efficiency is related to the fidelity of the lensing tracers and the instrumental noise in the CMB E-modes.
If we have a large scale structure field I($\nver$) that traces the lensing potential we can build a template of the lensing B-modes on the sky by a weighted convolution:
\be
\hat{B}^{\mathrm{lens}}(\bl) = \int \frac{d^2 \bl'}{(2 \pi)^2} W(\bl,\bl') f(\bl,\bl') E^N(\ublu) I(\bl - \bl'),
\ee
where $f(\bl,\bl')$ can be determined by minimizing the difference with the true $B^{\mathrm{lens}}(\bl)$ defined in \refeq{blens}. We include the instrumental noise in the CMB E-modes ($E^N$) that will also limit the ability to fully reconstruct the lensing B-modes.

The residual lensing B-modes due to an imperfect knowledge of the true E-mode and $\phi$ will be
\beqn
B^\mathrm{res}(\bl) &=&  B^\mathrm{lens}(\bl) - \hat{B}^\mathrm{lens}(\bl) =  \int \frac{d^2 \bl'}{(2 \pi)^2} W(\bl,\bl') \times \nonumber \\ &&\left( E(\ublu) \kappa(\bl - \bl' ) -  f(\bl,\bl') E^N(\ublu) I(\bl - \bl' ) \right).
\eeqn

The optimal weights $f(\bl,\bl')$, chosen such that the residual lensing B mode power is minimized, are \cite{sherwin:2015}:
\be
\label{eqn:fweight}
f(\bl,\bl') = \left(\frac{C^{EE}_{{l'}}}{C^{EE}_{{l'}}+N^{EE}_{{l'}}}\right)  \frac{C^{\kappa I}_{|\bl-\bl'|}}{C^{II}_{|\bl-\bl'|} }.
\ee
Here $C^{\kappa I}$ and $C^{II}$ are the cross-correlation spectrum of the tracer $I$ with the lensing convergence $\kappa$ and its autospectrum; they are described for each LSS field in \refsec{model}. The power spectrum of the E-modes noise $N^{EE}$ is the same as the B-modes one in \refeq{noise}.

With this choice of $f(\bl,\bl')$ we find that the residual power is:
\beqn
\label{eqn:Bres}
C_{{l}}^{BB,\mathrm{res}} &=&   \int \frac{d^2 \bl'}{(2 \pi)^2}  W^2 (\bl,\bl')
 C^{EE}_{l'} C^{\kappa \kappa}_{|\bl-\bl'|}  \\
&\times&  \left[1 - \left(\frac{C^{EE}_{{l'}} }{C^{EE}_{{l'}}+N^{EE}_{{l'}}}\right) \rho^2_{|\bl-\bl'|} \right] \nonumber
\eeqn
with
\be\label{eqn:rho1}
\rho_{\ell}^{2}= \frac{(\cl^{\kappa I})^{2}}{\cl^{\kappa \kappa}\cl^{I I}}.
\ee

\refeq{Bres} highlights the different factors that control the delensing efficiency.
The first part of the second term in the parenthesis consists of an inverse variance filter applied to the measured E-mode.
The smaller the noise in the E-modes ($N^{EE}$) is the closer this term is to a value of 1: a less noisy measurement improves the template of the lensing B-modes.
The second captures the difference between the reconstructed $\phi$ and the CMB lensing potential, and it directly relates the residual power after delensing with the cross-correlation coefficients with CMB lensing of the tracers used.
The larger the $\rho_{\ell}^{2}$ is for an LSS field, the more it is correlated with the lensing potential acting on the CMB photons. A higher correlation allows for a better reconstruction of $\phi$ and, as a consequence, of $B^{\rm{lens}}$ leading to a smaller residual power $C_{{l}}^{BB,\mathrm{res}}$.
We conclude this section showing in \reffig{bbplot} the expected residual lensing B-modes power spectrum for some of the tracers used in this work together with the primordial B-modes component and the instrumental noise for current and future experiments.

\begin{figure}[htbp]
\begin{center}

\includegraphics[scale=1.]{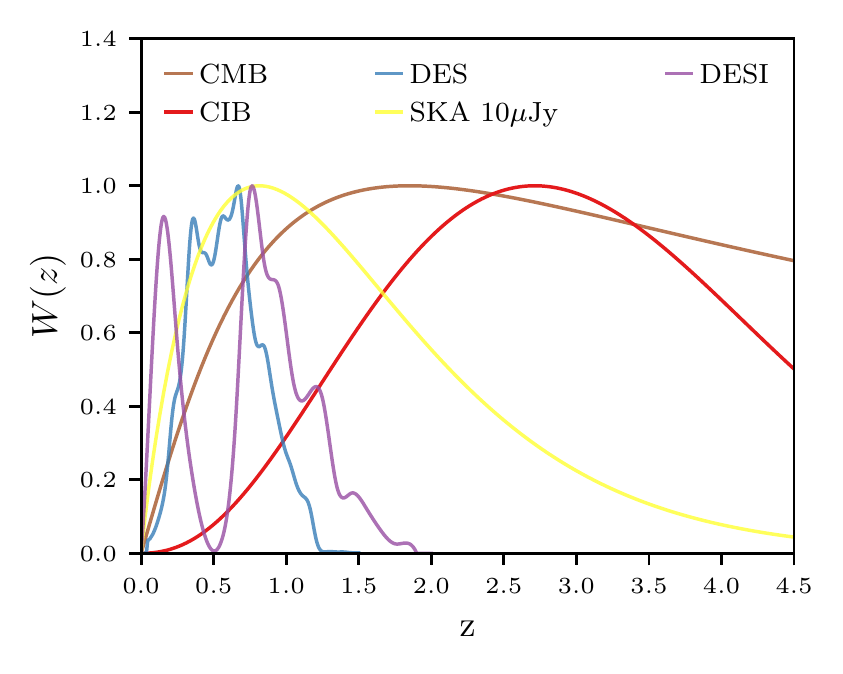}
\caption{Kernel Comparison: Comparison of the different kernels as a function of redshift for some of the tracers used in this analysis. The larger the overlap with the CMB lensing kernel the better the reconstruction of the lensing potential is thus leading to a higher delensing efficiency.
However, the efficiency is not exactly proportional to the overlapping area.
Firstly structure at different redshifts contribute differently to the generation of large-scale B-modes because of the geometric properties of the lensing kernel, and secondly, we are not taking the tracers noise into account here.
Note that for galaxy surveys the kernel corresponds to $b(z)\frac{dN}{dz}$ and not just the redshift distribution of galaxies.}
\label{fig:kernel}
\end{center}
\end{figure}


\begin{figure}[htbp]
\begin{center}
\includegraphics[scale=1.]{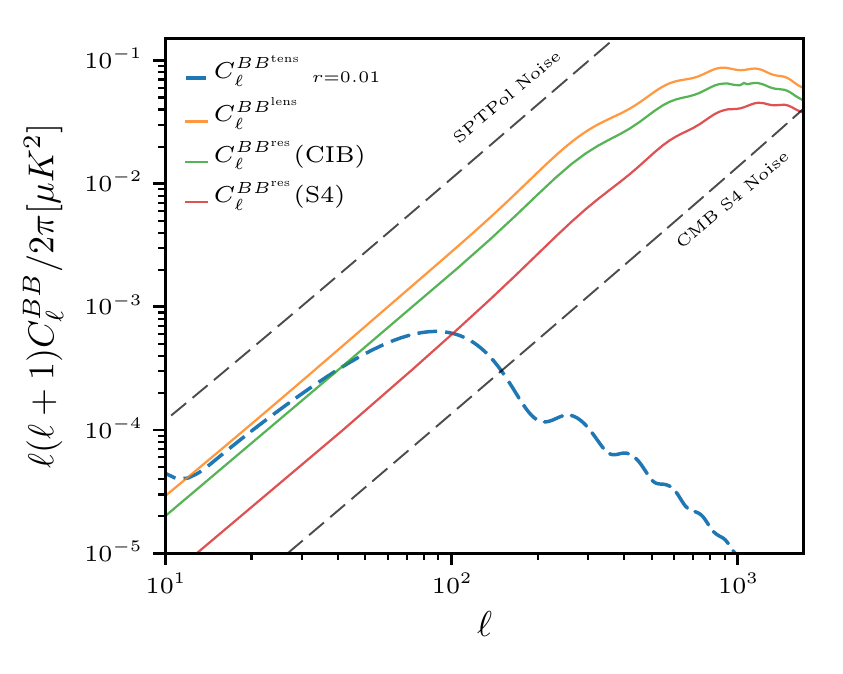}
\caption{Here we illustrate the effect of delensing on the B-mode power spectrum. The orange solid line corresponds to the fiducial lensing B-mode component of the signal while the dashed blue line corresponds to the inflationary one for the higher amplitude allowed by current experiments.
These are compared with the residual power left after delensing using some of the LSS tracers described in \refsec{model}.
The rapid improvement in the level of instrumental noise (dashed curves for SPTPol and CMB S4) will require a high delensing efficiency to exploit these experiments fully.}
\label{fig:bbplot}
\end{center}
\end{figure}

\subsection{Multiple tracers of the lensing potential}\label{sec:galcontrib}

In this section we extend the formalism to the case where multiple tracers are used to reconstruct the lensing potential.

We start by assuming that we have $n$ different tracers of the gravitational potentials $I_{i}$ with $i\in \{1,..,n\}$. We can optimally combine them to estimate $\phi$ or, in other words, to maximize the correlation factor $\rho$ with \cite{sherwin:2015}:
\beqn
I         &=&    \sum_{i}c^{i}I^{i} \nonumber \\
c_{i}     &=& (C_{II}^{-1})_{ij}C^{{\kappa I}^{j}}
\label{eqn:combined}
\eeqn
where $C_{II}$ is the covariance matrix of the LSS tracers. This is assumed to be gaussian and it is computed using \refeq{wkappa}.
The residual B-mode power can be derived from \refeq{Bres} using an ``effective'' correlation $\rho^{2}$ of these combined tracers with gravitational lensing:
\be
\rho^{2}_{\ell} = \sum_{i,j}\frac{C^{\kappa I^{i}}_{\ell}~(C^{II}_{\ell})^{-1}_{ij}~C^{\kappa I^{j}}_{\ell}}{C^{\kappa\kappa}_{\ell}}.
\label{eqn:rho-combined}
\ee

Note that the gain in adding a new tracer is not only proportional to its correlation with the CMB lensing, but it also depends on how much it is correlated with the already used set of tracers.
\reffig{kernel} show the different kernels as a function of redshift computed using the models and parameters described in \refsec{model}. The cross-correlation of a tracer with the CMB lensing is directly proportional to the overlap of their kernels.
It can be seen, for example, that the cosmic infrared background and 21 cm surveys probe the high redshift structures and, independent of the model assumed, they show a relatively good overlap with the CMB lensing kernel.
On the other end, galaxy clustering surveys can only reconstruct the low-z portion of the lensing kernel as can be seen from the LSST, DES and DESI curves.
However, given the low noise of these measurements and their small overlap with other probes, they can still play a major role in delensing even if their overlap with the CMB lensing potential is not optimal.

\subsection{Improving efficiency with tomographic binning}\label{sec:tomo-bin}
The delensing efficiency of galaxy surveys can be improved by taking into account redshift information.
When we weight a tracer with $\frac{C^{\kappa I}}{C^{II}}$ in \refeq{fweight} in order to maximize its ability to reconstruct the lensing potential we are only using redshift averaged information about the survey. However, as can be seen in \reffig{kernel}, the kernel overlap of a tracer with the CMB lensing varies as a function of redshift and the latter decreases steeply at low redshift. For this reason, the optimal approach is to weight galaxies at different redshift with different weights according to both their cross-correlation with $\kappa$ and their auto-spectrum.
We can see this with a simple example. Let's split a single survey I into two non-overlapping redshift bins $I_{1}$ and ${I_{2}}$ with $I=I_{1}+I_{2}$.
For the full survey the effective cross-correlation is equal to
\be
\rho_{\rm{full}}^{2}= \frac{(\cl^{\kappa I})^2}{\cl^{\kappa \kappa} \cl^{I I}}=\frac{(\cl^{\kappa I_{1}}+\cl^{\kappa I_{2}})^2}{\cl^{\kappa \kappa} (\cl^{I_{1} I_{1}}+\cl^{I_{2} I_{2}})}
\ee
while for the split survey it will be
\be
\rho_{\rm{split}}^{2}= \frac{(\cl^{\kappa I_{1}})^2}{\cl^{\kappa \kappa} \cl^{I_{1} I_{1}}}+\frac{(\cl^{\kappa I_{2}})^2}{\cl^{\kappa \kappa} \cl^{I_{2} I_{2}}}.
\ee
Now it can be show that $\rho_{\rm{split}}\ge \rho_{\rm{full}}$ since:
\be
\rho_{\rm{split}}^{2} - \rho_{\rm{full}}^{2}\propto  (\cl^{I_{1} I_{1}}\cl^{\kappa I_{2}}-\cl^{I_{2} I_{2}}\cl^{\kappa I_{1}})^2
\ee
Then $\rho^2$ is always larger in the tomographic case, and the two are equal only when $\frac{C^{\kappa I_{i}}}{C^{I_{i}I_{i}}}$ is the same for all the redshift bins in which case a single optimal weight is sufficient for the entire survey.
Binning will improve the efficiency of galaxy surveys, but it is not very effective for tracers with poor redshift information like the CIB or radio continuum surveys.

In this work, we bin both photometric and spectroscopic galaxy surveys by splitting the window function \refeq{wg} into different slices such as all the bins contain the same number of galaxies. For photometric surveys like DES and LSST, we assume a photometric redshift estimation Gaussianly distributed around the true value with an rms fluctuation $\sigma(z)$.

In that case the i$^{th}$ slice has a galaxy distribution \cite{hu:2004}:
\begin{eqnarray}
\label{eqn:zbin}
W_{i}(z) \propto b(z) {\frac{dN(z)}{dz}}
\Bigl[{\,\rm erfc}\left({\Delta(i-1) -z \frac{\sigma(z)}{\sqrt{2}}}\right)
\\ \nonumber - {\,\rm erfc}
\left({\Delta i  - \frac{z}{\sigma(z)}\sqrt{2}}\right)\Bigr].
\end{eqnarray}

\begin{figure}[htbp]
\begin{center}

\includegraphics[scale=1.]{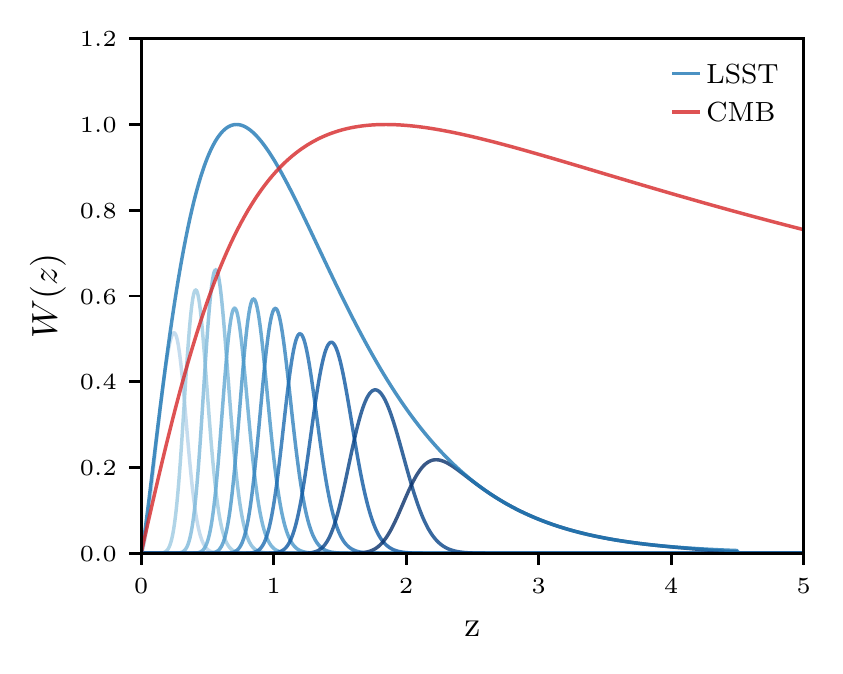}
\caption{Comparison of kernels of the 10 LSST tomographic bins together with the full LSST survey and the CMB lensing kernel.
Compared to a full survey approach, tomographic binning allows to optimally weight different bins according to their cross-correlation with the CMB lensing. This leads to a better delensing efficiency. }
\label{fig:kerneltomo}
\end{center}
\end{figure}

For photometric surveys, the maximum number of bins is dictated by the fact that the bin width cannot be smaller than the photo-z accuracy. Since we are not considering any possible photo-z bias, this is the only way photo-z uncertainties affect the delensing efficiency.
We used 10 and 4 photometric bins for LSST and DES respectively with a photo-z accuracy of $\sigma(z)_{\textrm{DES}} = 0.05(1+z)$ for DES and $\sigma(z)_{\textrm{LSST}} = 0.01(1+z)$ for LSST.

\begin{figure}[htbp]
\begin{center}
\includegraphics[scale=1.]{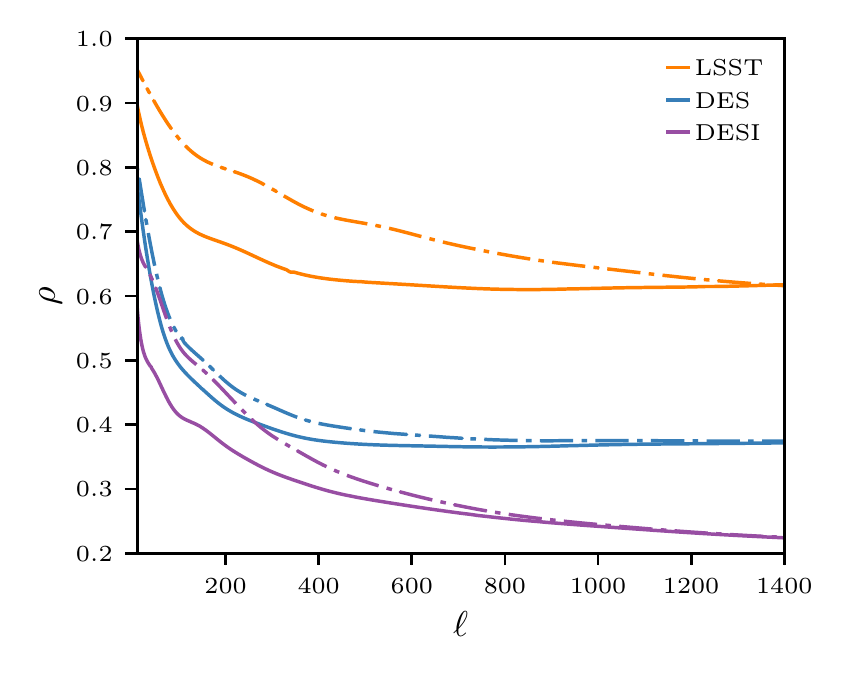}
\caption{Tomographic bins improve the cross-correlation of galaxy surveys with CMB lensing. Here we show the cross-correlation coefficient (\refeq{rho1}) as a function of angular scale for full surveys (solid lines) and tomographically binned surveys (dot-dashed line). For details about the binning, see \refsec{tomo-bin} and in particular \refeq{zbin}. }
\label{fig:rho-tomo}
\end{center}
\end{figure}

In spectroscopic surveys, there are no limitations in increasing the number of bins. We split DESI into 4 spectroscopic bins with no overlap among each others.
This number of bins is close to the saturation point where adding more bins does not improve delensing significantly while adding complexity to the analysis. As an example, we show in \reffig{kerneltomo} the 10 bins and the full LSST redshift distributions together with the CMB lensing kernel.
\reffig{rho-tomo} illustrate the improvement obtained by tomographic binning. In particular on large scales, binning can increase the value of $\rho$ by almost $30\%$ significantly improving the delensing efficiency of galaxy surveys.

\section{Parameter constraints improvement after delensing}
\label{sec:for}

In this section, we forecast the expected delensing efficiency and the relative importance of galaxy tracers for delensing in current and future experiments. We will use the Fisher information matrix to quantify the delensing efficiency as the improvement in the constraint of two inflationary parameters: the tensor to scalar ratio $r$ and the tensor tilt $n_{T}$.
We mainly focus on $r$ since these two parameters show similar improvement when delensing is moderate (does not remove more than $\sim 80\%$) and since, at least for single field inflation, only $r$ would be detectable by future experiments.

We assume a CMB experimental scenario composed of a high-resolution CMB experiment which defines the ability to internally reconstruct the lensing potential $\kappa$, together with a low-noise, low-resolution experiment that measures the B-modes that will be delensed and used to constrain the inflationary parameters.
As in \cite{abazajian:2016} and other works in the literature, we only combine the observations from the two experiments to obtain a single CMB E-mode map with optimally low noise.
We assume independent measurements for the lensing reconstruction and the B-modes measurement.
Depending on the experiments, it might be possible to optimally combined the two experiments and, once used for both lensing reconstruction and B-mode measurements, it can improve the level of delensing and the cosmological constraints.

Even if there is not a definite distinction between different experimental stages, we focus on three distinct scenarios: the ongoing or recently concluded stage (S2), a third generation stage (3G) of experiments in an advanced building phase or that have just started taking data and finally the futuristic CMB Stage 4.
s
\subsection{Fisher Information Matrix}\label{sec:fisher}
In the Fisher information matrix formalism \cite{10.2307/2342435}, the statistical uncertainty on a cosmological parameter $p$ can be obtained from the inverse of the Fisher matrix $F_{ij}$ as $\sigma(p) =\sqrt{(\mathbf{F})_{pp}^{-1}}$.
We constrain the inflationary parameters $p =\{r,n_{t}\}$ with a CMB B-mode spectrum measurement so the Fisher matrix is:
\beq\label{eqn:fisher}
F_{ij} = \sum^{\ell_{max}}_{\ell=\ell_{min}}\frac{1}{\sigma( C_{\ell}^{BB} )^2}\frac{\partial {C}_{\ell}^{BB}}{\partial p_{i}}\frac{\partial {C}_{\ell}^{BB}}{\partial p_{j}}
\eeq
where we assume a Gaussian covariance:
\beq\label{eqn:cova}
\sigma( C_{\ell}^{BB} )= \sqrt{ \frac{2}{(2 {l}+1) \fsky }} \left( C_{\ell}^{BB,r} + C_{{l}}^{BB,\mathrm{lens}}+N_{{l}}^{BB} \right).
\eeq
The B-modes noise spectrum is given by \cite{knox:1995}:
\beq \label{eqn:noise}
N_{\ell}^{BB} = \left({\Delta_P}/{T_{\mathrm{CMB}}}\right)^2  e^{ {l^2 \theta_\mathrm{FWHM}^2}/({8 \ln 2})}
\eeq
where $\theta_\mathrm{FWHM}$ is the full half width of the telescope beam, and $\Delta_P$ is the instrumental noise of the experiment.

It can be seen in \refeq{fisher} that removing the lensing contribution will improve parameter constraints.
The parameter uncertainties are inversely proportional to the covariance of the measurement. Since the lensing B-modes power $C_{{l}}^{BB,\mathrm{lens}}$ is a substantial component of the covariance, removing part of it through delensing will reduce the parameter statistical error.
In the following, we sometimes refer to the fraction of lensing B-mode power removed by delensing. This quantity is defined as: $\left< C_{\ell}^{BB,\mathrm{res}}/C_{{l}}^{BB,\mathrm{lens}}\right>_{4<\ell<100}.$

To quantify the improvement after delensing we introduce the error on a parameter $p$ constrained with \textit{delensed} spectra as $\sigma^{\rm{del}}(p)$.
Motivated by \cite{namikawa:2015} we assume the likelihood after delensing is well approximated by a multivariate gaussian. Then we compute $\sigma^{\rm{del}}(p)$ from \refeq{fisher} substituting $C_{{l}}^{BB,\mathrm{lens}}$ with the residual power after delensing $C_{{l}}^{BB,\mathrm{res}}$ in the covariance matrix defined in \refeq{cova}.
The improvement is then the ratio of the constraints before and after delensing: $\alpha_{r} = \sigma^{\rm{del}}(r)/\sigma(r)$ and $\alpha_{n_{t}} = \sigma^{\rm{del}}(n_{t})/\sigma(n_{t})$.
The ratio $\alpha_{r,n_{t}}$ is insensitive to the fraction of the sky observed by the experiment: while the absolute parameter constraints depend on $\fsky$ the relative improvement does not. For this reason, we do not quote $\fsky$ values for the CMB experiments investigated here.

\refeq{fisher} requires fiducial values for the parameters $p =\{r,n_{t}\}$. We explore different value for $r$ in \refsec{del_res}.
Given a fiducial value for $r$, the value of $n_{t}$ is fixed imposing the consistency relation $n_t=-r/8$ \cite{LIDDLE1992391}.
Furthermore, these results depend mildly on the choice of the pivot scale for the tensor and scalar primordial perturbation spectra.
In order to minimize the degeneracy between $r$ and $n_{t}$ \cite{abazajian:2016}, we choose as a pivot scale $k_{t}=0.01 $ Mpc$^{-1}$ .

In \refeq{fisher} we are making a few important assumptions. 
First, we are fixing all the cosmological parameters except $\{r,n_{t}\}$. Uncertainties in those will propagate to larger uncertainties in $\{r,n_{t}\}$. While neglecting this will lead to slightly optimistic absolute constraints, it has no significant impact on the estimate of the improvement due to delensing. For example, running the same pipeline with $10\%$ higher $\Omega h^{2}$, which is almost 40 times more than what Planck constraints allow \cite{planck-collaboration:2016b}, change the delensing improvement factors by at most $7-8\%$. A more reasonable 1\% higher $\Omega h^{2}$ corresponds to variations in the improvement factors of at most 1.5\%.

We are also neglecting an important contribution to the measured CMB B-modes: galactic polarized foregrounds.
The amplitude of these has been constrained in \cite{bicep2/keck-collaboration:2015, planck-collaboration:2015} and it strongly varies in different parts of the sky. Future experiments will use multi-band data to exploit the frequency dependence of these contaminants to remove them from the data. The amount of residual foregrounds depends both on uncertain foreground properties and experimental choices (see a review in \cite{abazajian:2016}).
For this reason, accurately treating foreground requires the use of simulations and the knowledge of several experimental details.
We decided to focus on an ideal situation assuming no foregrounds or perfect cleaning even if the importance of delensing will be slightly overestimated.

Furthermore, we are not considering the uncertainties on galaxy survey internal parameters such as biases, source distributions and photometric redshift uncertainties. These uncertainties can degrade inflationary constraints \cite{sherwin:2015,namikawa:2016a}.
However, as shown in \cite{sherwin:2015,namikawa:2016a}, these can be auto-calibrated i.e. they can be tightly constrained using galaxy survey auto- and cross-correlation spectra. We checked this for a few of the tracer combinations used here and find it particularly true once several tracers are jointly taken into account.
Finally, although we have assumed a Gaussian covariance $\sigma( C_{\ell}^{BB} )$ it actually has a non-Gaussian contribution \cite{motloch:2017,benoit-levy:2012}. This approximation is good enough to show the improvement due to delensing.
The approximations made here go in the direction of a slightly optimistic \textit{absolute} value for the parameters statistical uncertainties both for the standard and the delensed case. The delensing improvement is defined as the \textit{relative} value of these uncertainties $\alpha_{i} = \sigma^{\rm{del}}(i)/\sigma(i)$ and, at first order, it is not affected by these assumptions.

\subsection{Delensing with CMB Stage-2 and current LSSs}
\label{sec:del_res}


Recently delensing has been performed for the first time on data using both CIB maps \cite{larsen:2016a,manzotti:2017} and internally reconstructed CMB lensing maps \cite{carron:2017} as large scale structure tracers.
In this section we discuss the improvement that can be obtained by combining these and other currently Stage-2 available tracers (see \cite{yu:2017} for publicly accessible multi-tracers data-products).

On the CMB side, we will combine a BICEP-Keck like \textit{deep} CMB experiment with an overlapping \textit{higher resolution} experiment. For the deep experiment we assume an instrumental noise equal to $3\mu$K-arcmin in polarization ($\sqrt{2}$ lower in temperature), a beam of 30 arcmin and an angular scale range of $50<\ell<500$.
Here and in the following sections, we use all the scales measured by the instrument, even if the bulk of the signal to noise is at large scales $\ell<500$ \cite{dodelson:2014}.
We assume that the CMB lensing reconstruction is performed by the higher resolution CMB experiment.
We explore two possibilities.
First, we test the delensing efficiency with an internal reconstruction of CMB lensing performed by the Planck satellite \cite{planck-collaboration:2016a}. In this case, for the noise in the CMB lensing map in \refeq{noisekappa} we use the actual noise curves publicly available
\footnote{\url{https://wiki.cosmos.esa.int/planckpla2015/index.php/Specially\_processed\_maps}}. For consistency, we also combine a Planck-like E-mode map of $60\mu$K-arcmin in polarization with a beam equal to 7 arcmin that however adds almost nothing to the \textit{deep} CMB experiment.
Then, we test a second scenario with a ground-based experiment characterized by noise levels in polarization consistent with SPT-Pol \cite{henning:2017}: $9.4\mu$K-arcmin in polarization with a beam equal to 1.2 arcmin. In this case we set the angular scale range at $50<\ell<3000$. However, the results are quite robust against this choice.

We combine these CMB experiments with the CIB and current low redshift galaxy surveys like DES and WISE and we compute the improvement in the delensing efficiency.

Following \cite{yu:2017} we cut both the CIB and WISE at $\ell<100$ where they are contaminated by large Galactic dust residuals. 
In the future, with better theoretical foreground models or high signal to noise measurements, we might be able to characterize these fields (and their spectra $C_{\ell}^{\kappa I}$ and $C_{\ell}^{II}$) even at larger scales. Potential biases effect may be present like, for example, the correlation between the foreground contamination in galaxies and in the CMB polarization itself. 
However, we expect these biases to be small, and we might then be able to also use the $\ell<100$ component of these tracers to delens. However since the lensing contribution to B-modes from LSS modes at these scales is quite small, there is not too much to gain in term of delensing efficiency.

\begin{figure}[htbp]
\begin{center}
\includegraphics[scale=1.]{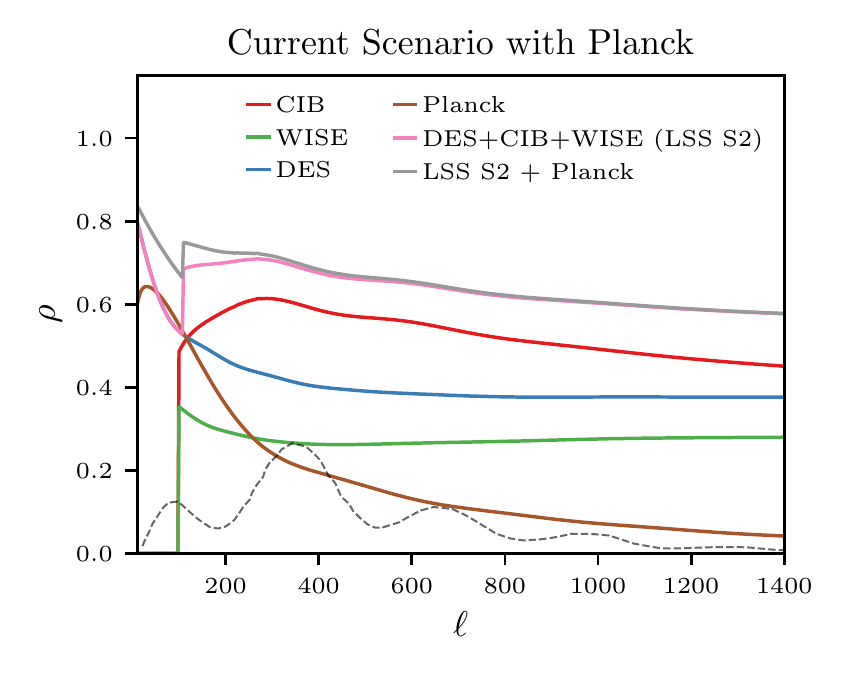}
\caption{Correlation factor $\rho$ with the CMB lensing potential as a function of the angular scale $\ell$ for completed and ongoing Stage-2 tracers.
Here we show both current galaxy survey and internally reconstructed CMB lensing potential.
The dashed curve highlights which scales contribute to the $\ell<100$ B-mode power and are thus most useful to delens. It corresponds to (with arbitrary scale) $C^{\kappa \kappa} \times \frac{\partial C^{BB}}{\partial{C^{\kappa \kappa}}}$}
\label{fig:corrnow-planck}
\end{center}
\end{figure}

\begin{figure}[htbp]
\begin{center}

\includegraphics[scale=1.]{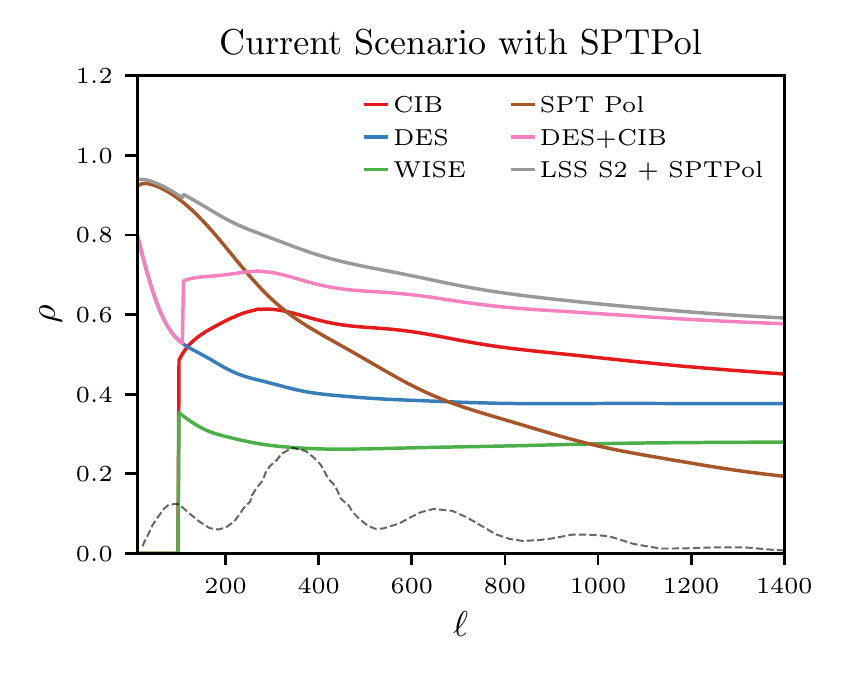}
\caption{Correlation factor $\rho$ with the CMB lensing potential as a function of the angular scale $\ell$ for completed and ongoing Stage-2 tracers.
This figure is the same as \reffig{corrnow-planck} but using SPTPol instead of Planck as the high-res CMB experiment performing the internal reconstruction.
}
\label{fig:corrnow}
\end{center}
\end{figure}

Optical surveys galaxies like DES are less affected by dust and can be used on larger scales.
The achievable correlation is shown in \reffig{corrnow-planck} for Planck and in \reffig{corrnow} for SPT-Pol.
In these figures, in order to show which scales in $\kappa$ contribute to the $\ell<100$ B-mode power, we include a dashed curve that corresponds to (with arbitrary scale) $<C_{\ell}^{\kappa \kappa} \times \frac{\partial C^{BB}_{\ell'}}{\partial{C^{\kappa \kappa}_{\ell}}}>_{\ell ' <100}$.

For Planck, the internal reconstruction is at most $\sim 65\%$ correlated at very large scales and then it falls rapidly to $40\%$ at $\ell=200$. Its correlation is comparable to the one of LSSs at almost all scales.
On the other end, the CMB lensing reconstruction from SPT-Pol will be more than $70\%$ correlated with the true field at $\ell<300$, and only then the correlation goes below the level of current LSS surveys.
In both cases the CMB lensing reconstruction correlates very well at low $\ell$ and it then falls rapidly at smaller scales because of the raise of the reconstruction noise.

\reffig{corrnow} and \reffig{corrnow-planck} show that the DES galaxies are effective tracers of the LSS and can, at least in the near future, be used to improve delensing for CMB experiments that overlap with it.
For example, DES delensing efficiency is higher than WISE first as a result of the lower level of noise and secondly because DES galaxies are located at slightly higher redshift so they better overlap with CMB lensing. As expected the improvement of the internal CMB reconstruction reduces the relative importance of the LSSs.
In particular it can be seen in \reffig{corrnow} that optical surveys will rapidly lose the role of filling in large-scale modes and they will start to supplement information at higher multipoles.

Using these correlation levels we can compute the residual B-mode power after delensing using \refeq{Bres}, and test the consequent improvement on parameter constraints with \refeq{fisher}.
We only report the improvements in $r$ but we still let $n_{t}$ to vary in our Fisher calculation even if it is not very degenerate with $r$.
We first compute the improvement with a fiducial value of $r_{\rm{fid}}=0$.
The results are summarized in \reftab{current} and \reftab{current-spt} for the Planck and SPTPol case respectively.

The lensing reconstruction noise in Planck is such that internal delensing can only remove $7\%$ of the power, improving the constraints on $r$ by a factor of 1.06 even in the ideal scenario of no instrumental noise in the B-modes. A slightly better performance is achieved with WISE. Once binned in redshift, DES can do better, removing a level of $17\%$ of power. However, it covers a smaller fraction of the sky than the two previous probes. As shown in previous works \cite{sherwin:2015}, the best tracer is the CIB with a $30\%$ reduction in power.
Nonetheless, it is worthwhile combining all the tracers. Indeed a multi tracer-approach can bring the removed power on the overlapping area from $30\%$ of the CIB alone to $42\%$ using all the LSSs (LSS-S2) and $45\%$ when Planck lensing reconstruction is added to the set.
However, if we consider a realistic level of noise in the measured B-modes, this level of residual power $C_{{l}}^{BB,\mathrm{res}}$ only leads to a 45\% improvement in $\sigma(r)$ for the null hypothesis case of $r=0$. Furthermore, as explained in \refsec{fisher}, this is only an upper limit and will be even smaller when, for example, non-perfect foreground cleaning is considered. Indeed current experiments are still significantly limited by foregrounds and instrumental noise together and not only by lensing.

If, on the other hand, we consider a high-resolution ground base experiment the internal lensing reconstruction improves significantly.
\reftab{current-spt} shows that, with an instrumental noise at the expected level for SPTPol, the CMB will soon be able to remove an amount of power $31\%$ at the level of the CIB. Once combined with a LSS tracer this will lead to removing $55\%$ of the power.
The improvement of 1.6 in $\sigma(r)$ that will follow, even when realistic noise is considered, is such that it will make delensing a needed step for CMB polarization data analysis.

\begin{table}
\caption{$\alpha(r)$: Improvements on $\sigma(r)$ due to delensing for completed and ongoing Stage-2 surveys and Planck lensing reconstruction.
LSS-S2 corresponds to the combination of all the available LSS tracers.
The values in parenthesis in the first column correspond to the fraction of lensing B-mode power removed using each LSS tracer.
It is defined as $\left<C_{\ell}^{BB,\mathrm{res}}/C_{l}^{BB,\mathrm{lens}}\right>_{4<\ell<100}$.
The values in the other columns correspond to ratio of the error before and after delensing for 3 cases: no instrumental noise in the B-mode measurement (but with instrumental noise in the E-mode used to delens) and with instrumental noise for two different values of $r$. The error $\sigma(r)$ is computed using \refeq{fisher} with $50<\ell<500$.}
  \vspace{0.2cm}
  \begin{tabular}{|c | c | c | c|}
\hline
Surveys & $\alpha_{r=0}, N^{B}_{\ell}=0$ & $\alpha_{r=0}$ & $\alpha_{r=0.07}$ \\ \hline \hline
WISE (8\%)& 1.08& 1.05 & 1.02 \\ \hline
DES (17\%) &1.20 & 1.13  & 1.05 \\ \hline
CIB (30\%) &1.44& 1.26  & 1.10\\ \hline
LSS-S2 (42\%) &1.75& 1.42  & 1.15\\ \hline
CMB Planck (7\%) &1.06 & 1.04  & 1.02 \\ \hline
LSS-S2 +CMB (45\%)& 1.81& 1.45& 1.16  \\ \hline
\end{tabular}
\label{tab:current}
\end{table}

We also test a scenario where primordial gravitational waves are present at their highest possible value of $r_{\rm{fid}}=0.07$ \cite{bicep2-collaboration:2016a}. As expected the importance of delensing itself is now reduced given that the lensing component constitutes a smaller portion of the total B-modes variance

 \begin{table}
\caption{$\alpha(r)$: Stage-2 improvements on $\sigma(r)$ due to delensing. We use the same LSS tracers as \reftab{current} but with the internal lensing reconstruction performed by SPTPol.}
  \vspace{0.2cm}
  \begin{tabular}{|c | c | c | c|}
\hline
Surveys &$\alpha_{r=0}, N^{B}_{\ell}=0$& $\alpha_{r=0}$ & $\alpha_{r=0.07}$ \\ \hline \hline
SPTPol (31\%) &1.43& 1.26  & 1.10 \\ \hline
LSS-S2 +SPTPol (55\%) & 2.2 & 1.6 & 1.20  \\ \hline

  \end{tabular}
  \label{tab:current-spt}
\end{table}


\subsection{CMB-S3 Era}\label{sec:S3}
CMB polarization measurements are rapidly improving. Indeed the next generation of ground-based telescopes has been already deployed, and data are currently being taken.
As we did for Stage-2, we model the CMB S3 as two different overlapping experiments.
For the deep experiment we assume an instrumental noise equal to $2\mu$K-arcmin in polarization and a beam of 30 arcmin.
The high resolution experiment will have a level of noise in polarization of $3\mu$K-arcmin and a 1 arcmin beam. This level of noise is also assumed for the internal noise reconstruction. For the CMB internal lensing reconstruction we use scales $50<\ell<3000$.
We use an angular scale range in the Fisher matrix of $50<\ell<800$ even if most of the high $\ell$ scale do not contribute to the constraints.
Furthermore, around 2019, DESI will start taking data. For this reason, we add DESI to the LSS tracers used in \refsec{del_res}.

\begin{figure}[htbp]
\begin{center}
\includegraphics[scale=1.]{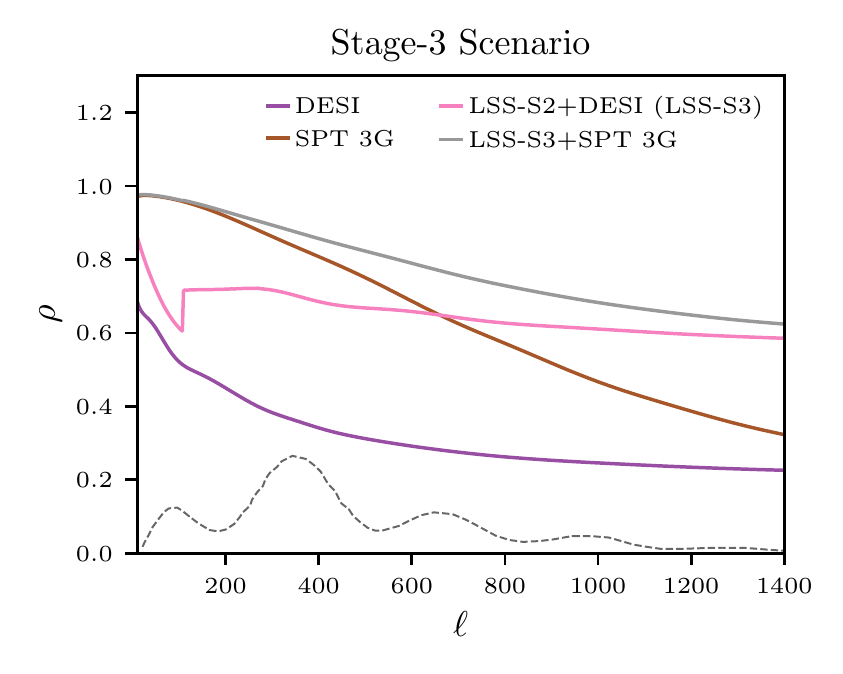}
\caption{Correlation factor $\rho$ with the CMB lensing potential as a function of the angular scale $\ell$. This figure is the same as \reffig{corrnow} but with a stage-3 CMB experiment performing the internal reconstruction and with DESI added to the stage-2 galaxy surveys (LSS-S2). }
\label{fig:corrS3}
\end{center}
\end{figure}

The correlation factor attainable using generation 3 experiments is shown in \reffig{corrS3}.
An interesting finding is that DESI will be less efficient (removing $11\%$ of the power) than a DES-like survey ($17\%$) despite the fact that it can probe slightly higher redshift. The reason is that, because of the broad CMB kernel, spectroscopic redshift accuracy is not needed and the lower shot noise in DES increases the delensing efficiency. Unfortunately, adding DESI will only bring the power removed using LSSs from 41 to 45\%.
Contrary to current S2 experiments, the CMB internal reconstruction will dominate the correlation with the lensing potential up to $\ell\simeq 600$.
For CMB S3, an iterative approach will improve the CMB lensing reconstruction compared to the quadratic estimator, even if not very significantly. For example, in this case, the removed power goes from $56\%$ to $51\%$ (quoted in \reftab{S3}) when using a quadratic estimator (QE) instead of the iterative one.

We summarize the improvement in the constraint on $r$ in \reftab{S3}.
Given the improvement in the noise of the high-resolution experiment, CMB alone will be able to improve constraint on $r$ by a factor of $\sim 2.3$ through internal delensing in the ideal case of no instrumental noise in the B-modes (but with instrumental noise in the E-mode used to delens).
Adding galaxy surveys will lead to a further improvement without additional effort.
Indeed galaxy surveys will still be able to remove an additional $12\%$ of power from $C_{\ell}^{BB,\mathrm{lens}}$ and to improve our constraint on the null hypothesis ($r=0, $) by $54\%$ in the noiseless case. Note that apart from the addition of DESI, the tracers used here are already available today. The ongoing effort can then lead to a significant improvement even for the next generation of experiments.
Finally, the combination of CMB and galaxy surveys will be able to improve the constrain on the null hypothesis of no primordial waves by a factor 2.3 or to improve the constraint of a possible detection (r=0.07) by 30\% even in the noisy case.
As expected the improvement on $n_{t}$ is similar to the one in $r$ since they are both proportional to the variance of the measured B-modes.
We tested this in the scenario $r=0.07$ where we additionally impose the consistency relation $n_{t}^{\rm{fid}}=-r^{\rm{fid}}/8$.
It is important to note that, independent of the level of delensing, for CMB 3G experiments the statistical error on $n_{t}$ will still be several times bigger than the fiducial value but it can still be tight enough to constrain more exotic inflationary models.

\begin{table}
\caption{$\alpha$: Improvements on $\sigma(r)$ due to delensing for S3 experiments. Here we define LSS-S3 as the combination of S2 LSS tracers (LSS-S2) and DESI. We also use a CMB S3 experiment for the lensing internal reconstruction. The error $\sigma(r)$ is computed using \refeq{fisher} with $50<\ell<800$, see \refsec{S3} for details.}
\hspace{2.2cm}
\begin{tabular}{|c | c | c | c|}
\hline
Surveys & $\alpha_{r=0}, N^{B}_{\ell}=0$ & $\alpha_{r=0}$ & $\alpha_{r=0.07}$ \\ \hline \hline
DESI (11\%) &1.13 & 1.1  & 1.04 \\ \hline
LSS-S3 (45\%) &1.82& 1.59  & 1.18\\

\hline
CMB 3G (QE) (51\%) &2.0& 1.68  & 1.20 \\
\hline
CMB 3G (56\%) &2.31& 1.9  & 1.23 \\ \hline
LSS-S3+CMB (68\%)& 3.17 & 2.3& 1.30  \\ \hline
\end{tabular}
\label{tab:S3}
\end{table}

\subsection{CMB-S4 Era}\label{sec:S4}
Having as a major goal the detection of inflationary B-modes, CMB data will continue improving even after Stage-3.
An ambitious program for a Stage-4 ground CMB experiment is in the planning phase \cite{abazajian:2016}.
Moreover, satellite and balloon CMB experiments have been proposed and have the potential to extend the accessible B-mode measurements to the largest scales. Given the unprecedented low level of noise of these experiment, delensing will be even more important than in previous generations.
Following the B-mode constraints section in \cite{abazajian:2016}, here we assume a CMB-S4 ground experiment composed of two different telescopes.
For the deep experiment, we assume an instrumental noise equal to $1\mu$K-arcmin in polarization and a beam of 15 arcmin. The high-resolution experiment will have a level of noise in polarization of $1.5\mu$K-arcmin and a 1 arcmin beam. Here we use the same angular scale range in the Fisher as the S3 case and the CMB internal lensing reconstruction is done using scales $50<\ell<4000$.

Furthermore, in the next decade, several next-generation LSS surveys will be online.
Even if these LSS surveys will observe an overlapping part of the sky and similar modes, because of their different strengths and weaknesses, it will still be important to combine them efficiently.
Here, as an example of future optical galaxy surveys, we add LSST \cite{lsst-science-collaboration:2009} to the CMB lensing tracers.
We do not show results for other surveys here but we test that other experiments like Euclid \cite{laureijs:2011} and WFIRST \cite{spergel:2013} have similar delensing performance.
We also consider a radio continuum survey modeled following the SKA specifications. Radio continuum observations of the 21 cm line are in their early stage, and several experimental and data-analysis challenges need to be overcome.
However, this technique has the potential to map the LSS with relatively low noise up to redshift $z=6$. Here we considered a detection threshold at 1 GHz (flux cut) of 10 $\mu$Jy which should be representative of SKA phase 1 (SKA1) \cite{namikawa:2016a}.
Given the importance of delensing for Stage-4 experiments, here we combine radio and optical survey to test their delensing efficiency.

\begin{figure}[htbp]
\begin{center}
\includegraphics[scale=1.]{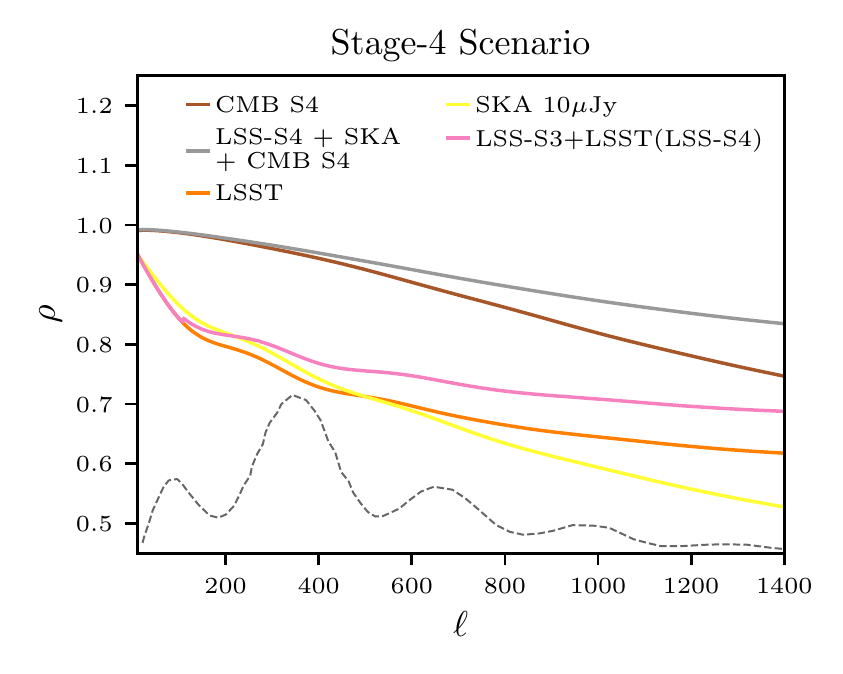}
\caption{Correlation factor $\rho$ with the CMB lensing potential as a function of the angular scale $\ell$. This figure is the same as \reffig{corrnow} but with a stage-4 CMB experiment performing the internal reconstruction and with LSST and SKA added to the stage-2 and stage-3 galaxy surveys (LSS-S3). }
\label{fig:corrs4_10}
\end{center}
\end{figure}

The correlation factor with CMB lensing for Stage-4 experiments is shown in \reffig{corrs4_10}.

A tomographically binned LSST-like experiment will be a very efficient CMB lensing tracer. Indeed it will be more than 70\% correlated with CMB lensing for all the scales $\ell<600$, a performance similar to a Stage-3 CMB internal reconstruction.
A very similar level of efficiency will be achieved by a SKA1-like experiment.
With SKA and LSST, for the first time we will have LSS tracers with a higher delensing efficiency than the CIB.
Despite the improvement in galaxy surveys, the CMB internal reconstruction will still be the main source of delensing. Having a perfect kernel overlap with the true lensing potential it will benefit from the very low level of noise of CMB S4 experiments.
\reffig{corrs4_10} shows that a CMB-S4 experiment will be able to internally reconstruct the lensing potential at more than 90\% up to $\ell=450$.
At this level, CMB S4 internal delensing will not be limited by noise but by small secondary effects like foregrounds contamination, filtered modes etc (see \cite{manzotti:2017}).

\reftab{S4} shows the improvement on the inflationary constraint given the cross-correlation factors described above.
LSST alone will be able to remove almost half of the B-mode lensing power. This will lead to an improvement in $\sigma(r)$ of a factor of 2. If we add all the other infrared and optical LSS tracers previously included (i.e. excluding SKA) the improvement factor increase to 2.4.

These gives an important opportunity: for sufficiently high values of $r$ it will be possible to confirm a detection of primordial gravitational waves using spectra delensed with LSS, completely independent of CMB internal data. 
Similarly to \cite{namikawa:2016a} we find that SKA phase-1 radio survey will remove 53\% of the lensing power and, once combined with optical and infrared surveys, this will allow the level of power removed using LSS (62\%) to surpass the level of the CMB S3 internal reconstruction.
Finally, as previously mentioned, CMB S4 will be the main source of delensing with the ability of removing up to more than 80\% of the B-modes power. 
Note that, at the noise levels of CMB-S4, it will be quite valuable to use an iterative algorithm for the lensing reconstruction. As shown in \reftab{S4}, a quadratic estimator will only be able to remove $73\%$ of power from $C^{BB}$ compared to the 83\% of the iterative approach.
This will bring a factor of $\sim 4$ improvement in the null test case even with realistic noise. Combining with LSS will only increase the removed power compared to CMB only by $\simeq 4.3\%$ but it will be highly beneficial to test the robustness of the final result.

For the case $r=0.07$, we find the improvement in $\sigma(n_{t})$ to be just slightly higher to the one in $\sigma(r)$ for all the CMB-S4 cases. This is reasonable because, given the level of noise in CMB-S4, the variance at angular scales $\ell<200$ are dominated by instrumental noise and not by the lensing component.
The conclusions about the relative importance of LSS tracers we find for $\sigma(r)$ are still valid for $\sigma(n_{t})$. Even if a test of consistency relation is not possible even for S4 experiment, delensing can still be useful also to constrain more exotic inflationary models.

\begin{table}
\caption{$\alpha$: Improvements on $\sigma(r)$ and $\sigma(n_{t})$ after delensing for S4 experiments. Here we define LSS-S4 as the combination of S3 LSS tracers (LSS-S3) and LSST. We also consider SKA-like radio-continuum surveys. We also use a CMB S4 experiment for the lensing internal reconstruction. The improvement for $\sigma(n_{t})$ is computed for fiducial values $r=0.07$ and $n_t=-r/8$. The error $\sigma(r)$ is computed using \refeq{fisher} with $50<\ell<800$ see \refsec{S4} for details.}
\begin{tabular}{|c | c | c | c| c|}
\hline
Surveys & $\alpha_{r=0}, N^{B}_{\ell}=0$ & $\alpha_{r=0}$ & $\alpha_{r=0.07}$ & $\alpha^{n_{t}}_{r=0.07}$ \\ \hline \hline
LSST (51\%) &2.1 & 1.9  & 1.21 & 1.47 \\ \hline
LSS-S4  (59\%) &2.4 & 2.1  & 1.25 & 1.57\\ \hline
SKA (10$\mu$Jy) (53\%) &2.1& 1.9  & 1.22& 1.46\\ \hline
LSS-S4 + SKA (62\%) &2.6 & 2.3  & 1.3& 1.6  \\ \hline
CMB S4 (QE) (73\%) & 3.7 & 2.9  & 1.35& 2.3 \\ \hline
CMB S4 (83\%) & 6.22 & 4.1  & 1.42& 1.9 \\ \hline
LSS-S4+CMB (86\%)& 7.7 & 4.7  & 1.46 & 2.46\\ \hline
\end{tabular}
\label{tab:S4}
\end{table}

\begin{figure*}[ht]
\includegraphics[scale=0.8]{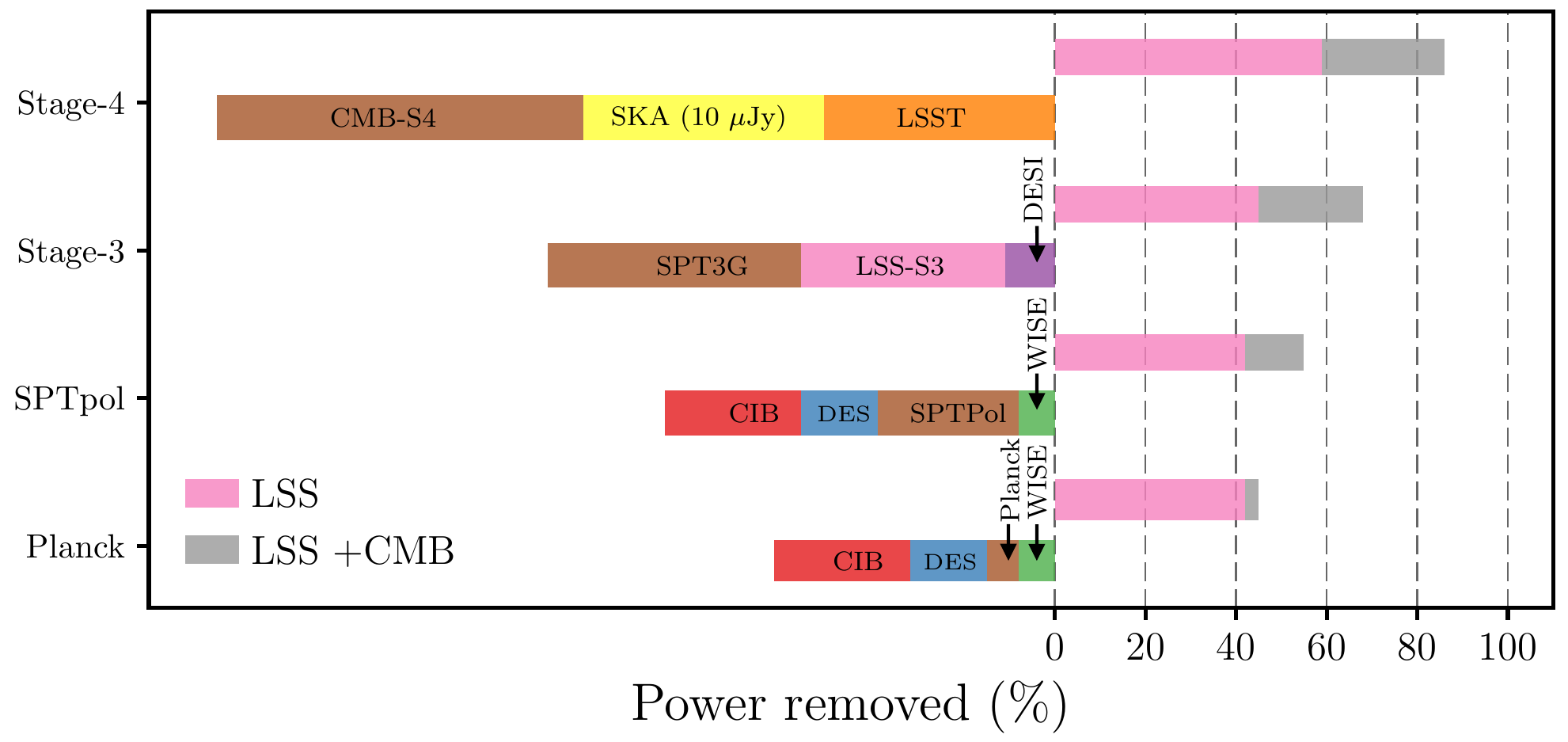}
\caption[]{A summary of the amount of lensing B-modes power removed by different tracers for different generations of experiments.

On the \textit{left} we show the contribution from each lensing tracers. The cross-correlation among tracers is not considered and for this reason, the sum of all the contributions can be bigger than one. The purpose is to highlight the relative importance of each tracer.

On the \textit{right} we show the power removed by LSS alone and final delensing efficiency once CMB internal delensing is added. Here the double-counting information is taken into account and the bars correspond to final delensing efficiency levels.  }
\label{fig:recap}
\end{figure*}


\section{Conclusions}
\label{sec:concl}
The ability to separate the lensing component of the CMB B-modes from a possible primordial inflationary signal (``delensing'') is necessary to test inflation using the next generation of CMB polarization experiments.
To delens, we need to accurately reconstruct the large scale structures that lens the CMB in order to marginalize the expected lensing B-modes components in the observed patch.
In this paper, we studied the potential impact of large-scale structure galaxy surveys in this important endeavor.
We focus on galaxy clustering here, leaving other probes like weak lensing, for future works. However, weak lensing is a lower redshift probes than clustering resulting in general in a worse delensing efficiency.
We find that, to improve the delensing efficiency of galaxy surveys, tomographic delensing will be very important: using several tomographic galaxy bins can improve the correlation of galaxies with CMB lensing by 10\%-30\% on a wide range of angular scales.

A summary of the delensing efficiency for different lensing tracers is shown in \reffig{recap}.
For ongoing experiments, we find that LSS tracers will be particularly beneficial.
An optical survey like DES is able to remove $17\%$ of the B-mode lensing power alone, and together with WISE($8\%$) and the CIB ($30\%$) will allow to remove $42\%$ of the power using only LSS surveys.
Depending on the CMB instrumental noise and the amount of galactic foreground cleaning, delensing using currently available LSS survey can correspond to a maximum improvement of $42\%$ in the constraint on the tensor to scalar ratio $r$ compared to the value before delensing.
In the future, the decreasing level of instrumental noise in CMB experiments will dramatically improve the internal reconstruction of the structures lensing the CMB.
The fraction of removed lensing B-modes will rapidly improve from the current expected levels for Planck ($7\%$) and SPTPol ($31\%$) to 3G (56\%) and CMB S4 (81\%) levels.
Indeed for Stage-3 experiments, the CMB internal reconstruction will be the main source of delensing. However, it will still be less efficient than galaxies at tracing the lensing potential at small scales $\ell>500$. For this reason, combining galaxy survey with the CMB will push the fraction of removed power from $56\%$ to $68\%$ for 3G.
Even for CMB S4, galaxy surveys will still play an important role.
For example, a tomographically binned LSST-like survey by itself will remove 51\% of the lensing power. This performance is lower than a S4 CMB internal reconstruction and comparable to a Stage-3 CMB.
However, it will allow us to probe the robustness against systematics of an eventual detection of primordial gravitational waves.
Indeed delensing with just CMB data will require a careful study of possible biases and systematic effects because we delens the same dataset (the CMB) that we also use to reconstruct the lensing potential \cite{carron:2017,sehgal:2017,namikawa:2017}.
For this reason, efficient galaxies tracers are not only useful in the short term but, in the future, will also play a role in testing internal biases and performing consistency tests.

Given the high level of correlation of future galaxy surveys and CMB lensing, it is also worth looking for potential delensing application besides the detection of inflationary B-modes.
For example, delensing with galaxy surveys allows us to selectively remove from the CMB only the gravitational effect coming from the non-linear structures in the low-redshift Universe.
Removing this component will reduce the level of non-linearities both in the CMB and in the CMB lensing power spectrum providing an alternative route to exploit all the measurable modes without having to model the non-linear component at small angular scales.

\section{Acknowledgments}
We thank T. Crawford, S. Dodelson, G. Fabbian, G. Holder, T. Namikawa, S. Passaglia, K. Story, A. Vieregg and K. Wu for useful discussions.
Furthermore we thank D. Alonso for his feedback on radio continuum surveys.

This work made use of computing resources and support provided by the Research Computing Center at the University of Chicago.
This work was partially supported by the Kavli Institute for Cosmological Physics at the University of Chicago through grants NSF PHY-1125897 and an endowment from the Kavli Foundation and its founder Fred Kavli.

\bibliographystyle{mnrasunsrt}
\bibliography{/Users/alessandromanzotti/Work/Astrophysics/latex_bib/cosmobib,/Users/alessandromanzotti/Work/Astrophysics/galaxies_delensing/Notes/spt,/Users/alessandromanzotti/Work/Astrophysics/galaxies_delensing/Notes/desi}

\end{document}